\documentclass[runningheads]{llncs}
\usepackage[T1]{fontenc}
\usepackage[utf8]{inputenc}
\usepackage{times}
\usepackage{soul}
\usepackage{url}
\usepackage[hidelinks]{hyperref}
\usepackage[small]{caption}
\usepackage{amsmath}

\usepackage{amsthm}
\usepackage{booktabs}
\usepackage{algorithm}
\usepackage{algorithmic}
\usepackage[switch]{lineno}
\usepackage{tikz}
\usepackage{makecell}
\usetikzlibrary{arrows.meta, positioning, shapes.geometric}
\usepackage{xcolor}
\usepackage{adjustbox}
\usepackage{longtable}
\usepackage{pdflscape}
\usepackage{microtype}
\usepackage{rotating}
\usepackage{graphicx}
\usepackage{array}

\newcommand{\twoline}[2]{%
  \begin{tabular}[c]{@{}c@{}}#1\\#2\end{tabular}%
}

\begin{document}
\title{Conversational AI for Perceived Cognitive Decline Screening in Older Adults}
%
%

\author{Nikita Gautam\inst{2} \and
Moana Sargent\inst{1} \and
Kaitlyn Sievers\inst{1} \and
Ashley Sylvara\inst{1,4} \and
Doina Caragea\inst{2} \and
Heather Bailey\inst{1} \and
Tianjun Sun\inst{1,3}}
\authorrunning{N. Gautam et al.}
%
\institute{Department of Psychological Sciences, Kansas State University \and
Department of Computer Science, Kansas State University \and
Department of Psychological Sciences, Rice University \and
Management Department, University of Florida}

\maketitle              

\begin{abstract}
Early identification of cognitive decline in older adults remains challenging because traditional neuropsychological assessments often require trained personnel, structured administration, and clinical access. Chatbot-based Intelligent Cognitive Assistants (ICAs) offer a scalable approach for collecting open-ended responses that may contain linguistic, semantic, and autobiographical cues related to cognitive health. This study evaluates whether ICA question-answer (Q/A) responses from older adults can support screening for participant-reported perceived cognitive decline and related cognitive, functional, and personality outcomes. We compare traditional machine learning models, frozen transformer embeddings, fine-tuned transformer models, and large language model (LLM)-based reasoning strategies, including Global Q/A analysis, Single Q/A confidence averaging, and Aggregated Q/A analysis. Among text-based methods, the Gemini 3.1 Flash Lite 1-example-per-class Aggregated Q/A configuration achieved the highest held-out test performance (0.742 accuracy, 0.739 balanced accuracy, 0.728 macro-$F_1$). Given the small test set ($n=31$) and the multiple configurations evaluated, these results are interpreted descriptively rather than as definitive model rankings.
Frozen clinical transformer embeddings with a random forest head were the strongest non-LLM text-based approach, while fine-tuned transformers showed limited generalization. Models using structured questionnaires, cognitive tasks, and personality measures achieved higher observed classification performance than models using ICA interview text. 
Regression analyses showed a modest predictive signal for various structured cognitive, functional, and personality-related measures
of cognitive decline, including AD8, FAQ, IQCODE, neuroticism, and conscientiousness. 
Exploratory question-level analyses identified several direct symptom-related and indirect autobiographical prompts with comparatively high test-set performance; however, effective sample sizes varied substantially across questions, and the question-level models were selected and ranked on the same held-out set.
These findings suggest that an ICA chatbot interview designed to measure cognitive functioning elicited measurable signals associated with perceived cognitive decline and may support scalable perceived cognitive decline screening research. Future research using larger, more clinically diverse samples and clinically validated outcomes is needed to support the use of these approaches in clinical settings.


\end{abstract}

\raggedbottom

\section{Introduction}

Early identification of cognitive change in older adults remains an important challenge for cognitive-health research and clinical decision support \cite{langa2014diagnosis}. Mild Cognitive Impairment (MCI) and other forms of early cognitive decline can be difficult to detect before they substantially affect everyday functioning, particularly because subtle changes in memory, attention, language, and functional ability often emerge gradually. Standard neuropsychological assessments and clinical screening instruments provide valuable information, but they typically require trained personnel, structured administration, and access to clinical resources \cite{apostolo2016predicting}. These requirements can limit scalability, particularly in rural and other underserved areas where the necessary clinical expertise and resources may not be readily available. As a result, there is growing interest in low-cost, accessible tools that can support early cognitive-risk screening while complementing, rather than replacing, formal clinical evaluation.

Conversational AI offers a promising direction for scalable cognitive health screening. Intelligent Cognitive Assistants (ICAs) and chatbot-based interview systems can administer semi-structured conversations and collect open-ended responses from older adults. Unlike fixed-response questionnaires, these interviews elicit naturalistic language that may contain linguistic, semantic, autobiographical, and behavioral cues related to memory, attention, verbal fluency, narrative organization, and functional self-reflection \cite{bucks1,bucks2,bucks3}. Prior ICA-based work has shown that older adults can engage meaningfully with chatbot-guided interviews and that these interactions can generate analyzable text for downstream modeling \cite{prototype}. However, it remains unclear how much cognitive-health signal is present in these responses and which computational methods are best suited to extract it.

The central question of this study is: \textit{to what extent can machine learning, transformer-based models, and LLMs extract perceived cognitive decline signals from chatbot-based question-answer responses?} Although ICA systems offer a promising platform for scalable cognitive-health screening, their practical value depends on whether computational models can recover reliable indicators from unstructured interview data. This question is especially important because several modeling strategies are now available, including traditional machine learning, frozen transformer embeddings, end-to-end transformer fine-tuning, LLM reasoning, structured-score-based baselines, and question-level analysis. A systematic comparison of these approaches is therefore needed to understand both the predictive potential and the limitations of ICA interview data.

To address this gap, we organize the study around three complementary evaluation tasks: 1) perceived cognitive decline classification and comparison with structured-score baselines, 2) continuous-outcome prediction from ICA interview text, and 3) question-level analysis of predictive signals. We compare multiple families of models under a consistent participant-level evaluation framework. Traditional linear and tree-based models serve as interpretable baselines, while frozen transformer encoders test whether pretrained language representations contain useful signal without task-specific fine-tuning. Fine-tuned transformer models evaluate whether end-to-end adaptation improves prediction from ICA text. We also evaluate LLM-based reasoning pipelines at three levels of textual granularity: 1) Global Q/A analysis of the full aggregated interview, 2) independent analysis of individual question-answer pairs, and 3) an aggregated reasoning strategy that combines intermediate outputs from question-level analysis. This design allows us to examine whether perceived cognitive decline signals are better captured by the full interview context, by specific high-value questions, or by structured aggregation across responses.

Rather than treating the ICA as a clinical MCI detector, we frame this study as an evaluation of whether open-ended chatbot interviews contain measurable signals associated with perceived cognitive decline, and whether LLM reasoning can extract these signals more effectively than conventional supervised models in a small-data setting. More specifically, we address the following four research questions:

\begin{itemize}
\small
    \item \textbf{RQ1:} How effectively can traditional machine learning, transformer-based, and LLM-based methods predict perceived cognitive decline from ICA chatbot interviews with older adults?

    \item \textbf{RQ2:} How does prediction from open-ended ICA interview text compare with prediction from structured questionnaire, cognitive-task, and personality measures?

    \item \textbf{RQ3:} To what extent can ICA interview text predict continuous cognitive, functional, and personality-related outcomes?
    
    \item \textbf{RQ4:} How are perceived cognitive decline signals distributed across full-interview and individual question responses, and which questions provide the strongest predictive evidence?
\end{itemize}

\section{Background and Related Work}
Mild Cognitive Impairment (MCI) is commonly described as an intermediate stage between normal cognitive aging and dementia, in which individuals exhibit measurable cognitive decline while often maintaining basic daily functioning \cite{petersen1999mci}. Early detection is important because individuals with MCI are at increased risk of developing Alzheimer's disease or related dementia, and earlier identification may support monitoring, planning, and timely intervention \cite{langa2014diagnosis,petersen2018practice}. 
However, traditional cognitive screening tools such as the Mini-Mental State Examination (MMSE) and Montreal Cognitive Assessment (MoCA), while relatively brief to administer, can be influenced by educational and cultural background and may have limited sensitivity to the subtle functional and linguistic changes associated with early mild cognitive impairment or Alzheimer’s disease \cite{nasreddine2005moca,apostolo2016predicting,edmonds2019early}. These limitations motivate scalable, accessible, and remotely deployable screening approaches that can complement formal clinical assessment.

Language-based assessment has become an important direction in cognitive health research because cognitive decline often affects memory retrieval, semantic organization, verbal fluency, coherence, and connected speech \cite{taler2008language,mueller2018connected}. Prior studies have shown that linguistic features extracted from speech or text can help distinguish cognitively healthy adults from individuals with MCI or dementia, especially when tasks require autobiographical recall, picture description, or open-ended narration \cite{aramaki2016vocabulary,fleming2014early,prototype}. Recent advances in natural language processing (NLP), transformer representations, and LLMs offer new opportunities to model these subtle linguistic and semantic cues from conversational data \cite{devlin2019bert,gemini,gpt3}. At the same time, language-based signals should be interpreted cautiously when outcomes reflect perceived cognitive change rather than clinically confirmed impairment.

Digital and AI-supported cognitive assessment tools, including ICAs, complement traditional screening by using chatbot-based interviews to collect rich, semi-structured narrative responses. Prior ICA research has shown that chatbot interviews can elicit meaningful text and support assessment of psychological attributes such as personality \cite{fan2023chatbot,sun2024development}. Recent work with older adults further suggests that they can meaningfully engage with ICA narrative interviews and generally perceive them as usable and accessible, although cognitive predictions from narrative text remain limited and depend on the design of cognitively relevant prompts \cite{prototype}.

Despite this progress, important gaps remain in the use of ICA interviews and LLMs for perceived cognitive decline screening. Recent work on LLMs suggests that models such as GPT, Gemini, and Claude can process unstructured natural language and support reasoning over complex clinical or psychological information \cite{gpt3,gemini,claude}. However, it remains unclear whether LLM-based reasoning provides advantages over traditional machine learning, frozen transformer embeddings, or fine-tuned transformer models in the context of chatbot-based cognitive-health screening.

It is also unclear whether the strongest predictive signal comes from full-interview representations, individual question-level responses, structured cognitive and questionnaire scores, or LLM-based aggregation across questions. This study addresses these gaps by evaluating traditional machine learning models, frozen transformer embeddings, fine-tuned transformer models, and LLM-based methods for predicting participant-reported perceived cognitive decline and related cognitive, functional, and personality outcomes from ICA chatbot interviews with older adults. In doing so, we examine whether LLMs can serve as a scalable reasoning layer for interpreting naturalistic ICA interview responses, while positioning the task as screening for perceived cognitive decline rather than clinical MCI diagnosis.

\section{ICA Design and Dataset}
A narrative interview study with older adults using an ICA chatbot was used to collect data for this study. The ICA conducts open-ended, text-based interviews and collects narrative responses that may contain linguistic, autobiographical, and cognitive signals relevant to cognitive functioning and aging.

\subsection{Participants and Data Source}
Participants were recruited from Prolific\footnote{\url{https://www.prolific.com}}, with eligibility restricted to adults aged 55 years or older, fluent in English, and based in the United States. The original study recruited 170 participants; after excluding those who failed attention checks or had incomplete data, the final analytic sample for our experiments comprised 153 participants.

\subsection{ICA Chatbot Interview Responses}
The ICA interview was administered through the Juji platform\footnote{\url{https://juji.io/docs/}} and followed a semi-structured narrative protocol. Participants responded to open-ended prompts about life history, school, work, family, friendships, hobbies, health motivations, and personal reflections. The ICA was not designed as a clinical diagnostic instrument. However, its initial interview questions were informed by items from established cognitive-decline and autobiographical or episodic-memory measures, including the CDR, MMSE, MoCA, ACE-III, and AMI. These items were adapted into open-ended conversational prompts requiring participants to retrieve autobiographical memories, organize their thoughts, maintain conversational context, and produce coherent written responses.

The interview contains both indirect autobiographical or narrative prompts and questions that more directly ask participants about perceived cognitive or functional changes. For example, some questions ask about conversational attention, spatial orientation, social engagement, driving difficulties, concentration, and everyday functioning, whereas others ask participants to recall occupational experiences, weddings, life events, or construct an open-ended story. This distinction is important because the primary classification outcome reflects participant-reported perceived cognitive decline. Consequently, performance on direct symptom-related questions may partly reflect semantic overlap with the target, whereas performance on indirect questions may reflect a broader mixture of linguistic, autobiographical, and cognitive information. The present analyses do not formally separate these sources of predictive signal.

\subsection{Outcome Variables}
\label{sec:outcome_variables}
We define both binary classification and continuous regression targets across three primary operational categories. 

The primary classification target, \textit{perceived\_decline}, was derived from a single-item subjective cognitive decline measure included in the intake questionnaire prior to the ICA interview. Participants were asked, ``\textit{Do you believe you're experiencing cognitive decline?}'' with binary response options: 1 (Yes) and 2 (No). Responses of 1 (Yes) were coded as perceived-decline (PD), whereas responses of 2 (No) were coded as no-perceived-decline (NPD). This measure was collected independently prior to the ICA conversation and was not derived from interview transcripts. No participants in the final analytic sample had missing values for this measure. The final analytic sample contained 100 participants in the no-perceived-decline class and 53 in the perceived-decline class. The held-out test set contained 20 participants labeled \texttt{NPD} and 11 labeled \texttt{PD}.

This output binary variable, as well as several continuous variables, are summarized in Table \ref{tab:outcome_variables}.

\begin{table}[ht]
\centering
\scriptsize
\setlength{\tabcolsep}{6pt}
\begin{tabular}{p{0.28\linewidth} p{0.20\linewidth} p{0.46\linewidth}}
\toprule
Category & Variable / Metric & Description \\
\midrule
Primary Classification & perceived\_decline & Self-reported binary indicator of cognitive decline. \\
\midrule
Structured Questionnaire Measures
& FAQ\_score
& Functional Activities Questionnaire; instrumental daily living tasks (Higher = greater difficulty). \\

& AD8\_score
& AD8 Dementia Screening Interview; self-reported change in memory and judgment (Higher = more reported decline). \\

& IQCODE\_score
& Informant Questionnaire on Cognitive Decline; 10-year cognitive trajectory (Higher = worse decline). \\
\midrule
Objective Cognitive Functioning & lc\_proportion\_correct & Letter Comparison task accuracy over a timed 60-second trial interval (Higher = faster processing speed). \\
\midrule
Personality Traits & BFI-2 traits\cite{soto2017next} & Big Five Inventory; metrics for extraversion, agreeableness, conscientiousness, neuroticism, openness. \\
\bottomrule
\end{tabular}
\caption{Summary of target participant-level outcome variables.}
\label{tab:outcome_variables}
\end{table}

\section{Experimental Setup}

\begin{figure}[htbp]
    \centering
    \includegraphics[width=0.85\linewidth]{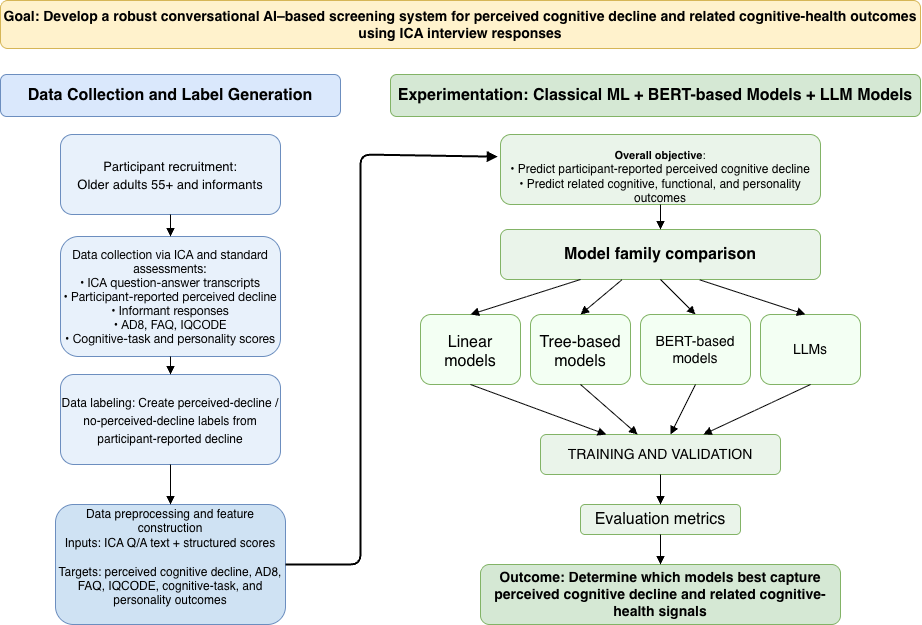}
    \caption{Overall Methodology}
    \label{fig:methodology}
\end{figure}

This section details the framework used to evaluate whether ICA interview responses can predict participant-reported perceived cognitive decline and related outcomes. An overview of the framework is shown in Fig. \ref{fig:methodology}. We compare raw text, structured questionnaire data, and cognitive-task data, along with various modeling strategies. 

\subsection{Prediction Tasks}
We define three primary evaluation tasks to answer our four research questions:
\begin{itemize}
\small
\item \textbf{Task 1: Perceived cognitive decline classification and input comparison}: Predicts participant-reported perceived cognitive decline from ICA interview text and compares the best text-based model with models trained on structured questionnaire, cognitive-task, and personality measures.

\item \textbf{Task 2: Cognitive, functional, and behavioral outcome regression using Q/A responses}: Evaluates whether ICA interview responses can predict continuous participant-level outcomes collected through separate questionnaire and cognitive-task measures, including AD8, FAQ, IQCODE, Letter Comparison proportion correct, and Big Five Inventory traits.
\item \textbf{Task 3: Per-question analysis for interpretable evidence}: Evaluates each ICA question independently to identify which prompts provide the strongest evidence for participant-reported perceived decline.
\end{itemize}

\subsection{Data Splitting and Validation Strategy}

The dataset contains 153 participants with matched ICA interviews and outcome measures. All splits are strictly partitioned at the participant level to prevent data leakage. We use a fixed 80/20 stratified split, yielding 122 participants for development and 31 for held-out testing.
Because the final test set contains only 31 participants, held-out performance should be interpreted as an independent but sample-limited estimate of generalization. A difference of only one or two correctly classified participants can noticeably change accuracy and related metrics, so small differences among the strongest models should not be interpreted as precise rankings.

Within the development set, model selection and hyperparameter tuning (via Optuna \cite{optuna}) are performed using nested stratified cross-validation (5-fold outer loop for generalization; 4-fold inner loop for tuning). For classification, the optimization objective is macro-$F_1$. A fixed random seed of 42 was used to enable reproducibility.

\subsection{Modeling Approaches}
We evaluate four distinct modeling families:

\subsubsection{Traditional ML Baselines}
Linear and tree-based models, e.g., Logistic/Linear Regression, Support Vector Machines (SVM), Support Vector Regression (SVR), and Random Forests (RF), trained on inputs or standard features.

\subsubsection{Frozen Transformer Embeddings}
Pretrained transformer encoders are utilized as fixed feature extractors. Lightweight linear and Random Forest heads are trained on the resulting participant-level embeddings.

\subsubsection{Fine-Tuned Transformers}
End-to-end fine-tuning of domain-general and clinical language models (BERT-base/large, BioClinicalBERT, ClinicalBERT, Clinical-Longformer, and BioClinical ModernBERT). Short-context models use a hierarchical architecture with masked mean pooling across Q/A pairs, while long-context models ingest full concatenated interviews. For short-context models, including BERT and ClinicalBERT variants, each Q/A pair is encoded separately with a maximum length of 128 tokens and participant-level representations are obtained by masked mean pooling over available question embeddings. A task-specific prediction head is then trained on the pooled representation. Long-context models, including Clinical-Longformer and BioClinical ModernBERT, ingest the full concatenated interview.

\subsubsection{LLM-based Classification}
We evaluate LLMs for participant-reported perceived cognitive decline classification using a prompt-optimized reasoning pipeline. Unlike the supervised models, the LLM setting frames the task as a research classification problem rather than a clinical diagnostic task. The models are explicitly instructed not to provide a clinical diagnosis, but to classify participants based only on the provided ICA Q/A responses. The LLM experiments use the same development and held-out test split as the supervised models. Development data are used for prompt optimization and few-shot example selection, whereas the held-out test set is reserved exclusively for final evaluation.

Prompt optimization was performed entirely within the 122-participant development partition. For the LLM experiments, this partition was further divided using seed 123 into a prompt-optimization training pool ($n=79$) and validation pool ($n=43$); the held-out 31-participant test set was not used during prompt development. During each optimization iteration, up to 24 participants were sampled from each development pool. Optimization was performed under the zero-shot Global Q/A condition. Misclassified training examples, up to eight per iteration, were supplied to an optimizer LLM to propose revisions to five prompt modules: task definition, reasoning policy, decision rules, few-shot guidance, and output format.

Two optimization iterations were attempted for each task model. Candidate prompts were evaluated on a sample of development-validation participants, and the prompt with the highest validation macro-$F_1$ was retained, using balanced accuracy and standard accuracy as successive tie-breakers. The selected prompt was fixed before the held-out evaluation.

For the final evaluation, we compare zero-shot prompting with one-example-per-class prompting. The zero-shot condition contains no labeled demonstrations. In the one-example-per-class condition, one \texttt{PD} and one \texttt{NPD} participant from the development data are included as demonstrations. For the one-example-per-class condition, one participant from each outcome class was selected from the development partition using a fixed few-shot sampling seed of 42, after excluding all held-out test participant identifiers. This seeded procedure produced the same two development examples, which were then reused across GPT-4o, Claude, and Gemini for all held-out evaluations.

Each LLM response is constrained to a structured JSON output containing the predicted class, class probabilities, evidence strength, and a brief explanation supporting the prediction. We evaluate three inference strategies. The \emph{Global Q/A} strategy provides all available Q/A pairs for a participant in a single prompt and requests one participant-level prediction. The \emph{Single Q/A Confidence Average} strategy prompts the LLM separately for each available Q/A pair and averages the LLM-reported class confidence scores across questions to obtain the participant-level prediction. Finally, the \emph{LLM-based Aggregated Q/A} strategy first obtains the individual Single Q/A analyses and then uses a second aggregation prompt to combine the question-level predictions, confidence scores, evidence strengths, and brief explanations into a final participant-level decision. These three strategies allow us to compare reasoning across the complete interview context, independent question-level reasoning, and an additional LLM-based meta-aggregation step.

\subsection{Evaluation Metrics}
Classification performance is measured using accuracy, balanced accuracy, macro-$F_1$, and weighted-$F_1$, with priority given to balanced accuracy and macro-$F_1$ due to class imbalance. Regression models are assessed via the coefficient of determination ($R^2$). Beyond predictive performance, models are evaluated qualitatively for their ability to surface interpretable, question-level cognitive health signals.

To quantify test-set performance uncertainty ($n=31$), we conducted a participant-level bootstrap (10,000 resamples with replacement, seed 42) to derive empirical 95\% percentile confidence intervals for balanced accuracy and macro-$F_1$. For accuracy, we report analytical Wilson 95\% score intervals.

\subsection{Hyperparameter Selection and Final Testing}
Hyperparameters (e.g., learning rate, regularization strength, tree depth) are tuned exclusively in the inner cross-validation folds. Selection relies on macro-$F_1$ for classification. Final evaluation metrics are reported on the independent, 31-participant held-out test set.

\subsection{LLM Reproducibility Details}

We evaluated GPT-4o (\texttt{gpt-4o}), Claude Opus 4.6 (\texttt{claude-opus-4-6}), and Gemini 3.1 Flash Lite (\texttt{gemini-3.1-flash-lite-preview}). The LLM APIs were accessed between April 2026 and June 2026. Final task-model inference used temperature $=0.0$, a maximum output length of 1,200 tokens, and up to three retries. No additional optional decoding or sampling parameters were specified in the API requests beyond these settings; in particular, parameters such as top-$p$, top-$k$, frequency or presence penalties, and generation seeds were not explicitly set. Prompts included task instructions, reasoning and decision guidance, few-shot instructions, and a structured output format. Complete prompts, configurations, few-shot examples, and saved outputs are available at \href{https://github.com/NikitaGautam/CogHealth-Year1}{GitHub}\footnote{\url{https://github.com/NikitaGautam/CogHealth-Year1}}.

Prompt optimization was performed separately for each model using only development data, divided into training and validation pools. Two optimization iterations were attempted using full interviews, with up to 24 participants sampled per pool and up to eight training errors used to guide revisions. The optimizer models were \texttt{gpt-4o-mini}, \texttt{claude-3-5-haiku-latest}, and \texttt{gemini-1.5-flash}, respectively, with temperature $=0.2$ and a 900-token output limit. Archived artifacts show that GPT-4o retained a first-iteration prompt revision, whereas the final Claude and Gemini prompts remained identical to their initial versions.

The prompt with the highest validation macro-$F_1$ was selected, with balanced accuracy and accuracy used as tie-breakers, and was fixed before held-out testing. Zero-shot and one-example-per-class settings were then evaluated using examples selected only from development data; test results were not used for prompt revision or example selection.

\section{Results and Analysis}
\subsection{Perceived Cognitive Decline Classification Using All Q/A Responses}

\begin{table}[htbp]
\centering
\scriptsize
\setlength{\tabcolsep}{5pt}
\begin{adjustbox}{max width=\textwidth}
\begin{tabular}{llrrrr}
\toprule
Setting & Model & Acc. & Bal. Acc. & Macro F1 & Weighted F1 \\
\midrule

\multicolumn{6}{l}{\textbf{Frozen Transformer Embeddings + Random Forest Head}} \\
Frozen + RF & medicalai\_ClinicalBERT
& 0.645 & 0.582 & 0.583 & 0.630 \\
Frozen + RF & distilbert-base-cased
& 0.645 & 0.561 & 0.554 & 0.613 \\

\midrule
\multicolumn{6}{l}{\textbf{Frozen Transformer Embeddings + Linear Classification Head}} \\
Frozen + Linear & BioClinical-ModernBERT-large
& 0.677 & 0.545 & 0.483 & 0.575 \\
Frozen + Linear & Clinical-Longformer
& 0.645 & 0.500 & 0.392 & 0.506 \\

\midrule
\multicolumn{6}{l}{\textbf{Linear and Tree-Based Baselines}} \\
Traditional ML & random\_forest
& 0.613 & 0.577 & 0.577 & 0.613 \\
Traditional ML & logistic\_regression
& 0.613 & 0.536 & 0.530 & 0.587 \\

\midrule
\multicolumn{6}{l}{\textbf{Fine-Tuned Transformer Models}} \\
Fine-tuning & bert-base
& 0.548 & 0.445 & 0.410 & 0.493 \\
Fine-tuning & medicalai\_ClinicalBERT
& 0.645 & 0.500 & 0.392 & 0.506 \\

\midrule
\multicolumn{6}{l}{\textbf{LLM-Based Reasoning Methods}} \\
LLM & Claude Opus 4.6, 0-shot, Global Q/A
& 0.677 & 0.668 & 0.660 & 0.682 \\

LLM & Gemini 3.1 Flash Lite, 0-shot, Global Q/A
& 0.742 & 0.718 & 0.718 & 0.742 \\

LLM & Gemini 3.1 Flash Lite, 1-per-class, Aggregated Q/A
& \textbf{0.742} & \textbf{0.739} & \textbf{0.728}
& \textbf{0.746} \\

LLM & GPT-4o, 0-shot, Global Q/A
& 0.710 & 0.714 & 0.698 & 0.715 \\

\bottomrule
\end{tabular}
\end{adjustbox}
\caption{Core all-Q/A perceived cognitive decline classification results across frozen transformer embeddings, traditional machine learning baselines, fine-tuned transformer models, and LLM-based reasoning methods. Representative configurations are reported for each model family. Given the 31-participant held-out test set, these differences should be interpreted cautiously. For example, one additional correct or incorrect participant changes accuracy by approximately 0.032. The results therefore support broad comparisons among modeling families more strongly than fine-grained claims that one high-performing LLM configuration is definitively better than another.}
\label{tab:results-aim1}
\end{table}

Table~\ref{tab:results-aim1} summarizes the classification results. Among the held-out ICA-text configurations, the highest numerically observed balanced accuracy and macro-$F_1$ were achieved by Gemini 3.1 Flash Lite with one-example-per-class prompting and LLM-based Aggregated Q/A analysis, reaching 0.742 accuracy, 0.739 balanced accuracy, 0.728 macro-$F_1$, and 0.746 weighted-$F_1$. This configuration should be interpreted as a descriptive maximum rather than a test-independent model-selection result. We evaluated 18 LLM configurations on the same held-out test set (three task models $\times$ two shot conditions $\times$ three inference strategies). Across these configurations, accuracy ranged from 0.484 to 0.742, balanced accuracy from 0.441 to 0.739, and macro-$F_1$ from 0.380 to 0.728. Consequently, the highest-observed configuration may benefit from post hoc selection among multiple test-set evaluations.

Second, among the reported non-LLM text-based approaches, frozen transformer embeddings provided stronger held-out performance than the fine-tuned transformer configurations. The strongest non-LLM model was medicalai\_ClinicalBERT with a random forest head, which achieved 0.645 accuracy, 0.582 balanced accuracy, and 0.583 macro-F1. This suggests that pretrained clinical language representations contain useful signal, especially when paired with a non-linear classifier. Third, fine-tuned transformer models showed limited generalization. Most fine-tuned models achieved moderate accuracy, with medicalai\_ClinicalBERT having 0.500 balanced accuracy and 0.392 macro-F1, indicating near-majority-class prediction behavior. This pattern likely reflects the small sample size, class imbalance, and difficulty of end-to-end fine-tuning with limited participant-level data. Overall, these results support LLM-based reasoning and frozen embedding approaches as more suitable than full fine-tuning for this small-data ICA interview setting.

\subsection{Comparison with Structured-Score Baselines}

\begin{table}[t]
\centering
\scriptsize
\setlength{\tabcolsep}{6pt}
\begin{tabular}{llccc}
\toprule
Input & Reported configuration & Acc. & Bal. Acc. & Macro-F1 \\
\midrule
ICA interview text
& Gemini, Aggregated Q/A
& 0.742 & 0.739 & 0.728 \\

Structured scores
& Extra Trees
& \textbf{0.871} & \textbf{0.859} & \textbf{0.859} \\
\bottomrule
\end{tabular}
\caption{Numerically highest held-out classification results using ICA text and structured questionnaire, cognitive-task, and personality measures.}
\label{tab:text-structured-comparison}
\end{table}

Table~\ref{tab:text-structured-comparison} compares the numerically highest ICA-text and structured-score results. Extra Trees trained on structured measures achieved 0.871 accuracy, 0.859 balanced accuracy, and 0.859 macro-$F_1$, numerically exceeding the selected ICA-text configuration on the held-out set. Its accuracy is 27/31 correct predictions. This comparison should nevertheless be interpreted as a reference baseline because measures such as AD8 and IQCODE assess constructs related to perceived cognitive and functional change and therefore are not fully independent of the primary classification target.

\subsection{Uncertainty Analysis of Held-Out Classification Results}

\begin{table}[t]
\centering
\scriptsize
\setlength{\tabcolsep}{4pt}
\begin{tabular}{lccc}
\toprule
Model 
& \makecell[c]{Accuracy\\{[95\% CI]}} 
& \makecell[c]{Bal. Acc.\\{[95\% CI]}} 
& \makecell[c]{Macro-$F_1$\\{[95\% CI]}} \\
\midrule
Extra Trees (structured) 
& .871 [.711, .949] 
& .859 [.710, .977] 
& .859 [.705, .966] \\

Gemini, 1/class, Aggregated Q/A 
& .742 [.568, .863] 
& .739 [.566, .895] 
& .728 [.551, .871] \\

Gemini, 0-shot, Global Q/A 
& .742 [.568, .863] 
& .718 [.543, .879] 
& .718 [.532, .868] \\

GPT-4o, 0-shot, Global Q/A 
& .710 [.534, .839] 
& .714 [.538, .875] 
& .698 [.516, .852] \\

Claude, 0-shot, Global Q/A 
& .677 [.501, .814] 
& .668 [.485, .838] 
& .660 [.479, .820] \\

ClinicalBERT + RF 
& .645 [.469, .789] 
& .582 [.411, .755] 
& .583 [.388, .758] \\
\bottomrule
\end{tabular}
\caption{Held-out uncertainty estimates ($n=31$). Accuracy uses Wilson 95\% score intervals; balanced accuracy and macro-$F_1$ use 95\% participant-level bootstrap intervals (10,000 resamples).}
\label{tab:classification-bootstrap-ci}
\end{table}

Table~\ref{tab:classification-bootstrap-ci} shows that uncertainty is substantial because the held-out test set contains only 31 participants. Although Gemini Aggregated Q/A produced the numerically highest balanced accuracy and macro-$F_1$ among the ICA-text approaches, its bootstrap confidence intervals overlap substantially with those of Gemini Global Q/A, GPT-4o, Claude, and the ClinicalBERT + RF baseline. Thus, the observed differences among the text-based models should not be interpreted as establishing a definitive ranking. The structured-score Extra Trees model showed numerically higher performance, although its confidence intervals were also relatively broad. These results support broad differences in observed performance while emphasizing the uncertainty of fine-grained model comparisons in this small held-out sample. 

\subsection{Continuous Outcome Prediction Using All Q/A Responses}
\label{sec:results_allqa_regression}

Table \ref{tab:results-aim3} reports selected all-Q/A regression results using $R^2$, with representative and best-performing configurations shown for each model family. Predicting continuous outcomes from ICA responses was substantially more difficult than perceived cognitive decline classification, with many models producing $R^2$ values near zero or below the mean-prediction baseline. Nevertheless, modest predictive signals were observed for AD8, FAQ, IQCODE, neuroticism, and conscientiousness. Frozen transformer embeddings with linear regression heads generally provided the strongest and most consistent results. The best $R^2$ values were obtained for neuroticism ($0.218$), conscientiousness ($0.209$), IQCODE ($0.189$), AD8 ($0.183$), and FAQ ($0.174$). Letter Comparison performance was poorly predicted, with a maximum $R^2$ of only $0.057$. Traditional regressors and end-to-end fine-tuned transformers showed limited or unstable generalization. Overall, the ICA interview text contains a modest signal for selected outcomes, but does not reliably predict most continuous measures in this small-sample setting.

\begin{table*}[h]
\centering
\scriptsize
\setlength{\tabcolsep}{3.6pt}
\renewcommand{\arraystretch}{1.18}

\begin{adjustbox}{max width=\textwidth}
\begin{tabular}{@{}>{\raggedright\arraybackslash}p{2.35cm}
                >{\raggedright\arraybackslash}p{3.15cm}
                rrrrrrrrr@{}}
\toprule
Setting & Model & AD8 & \twoline{BFI}{Agree} & \twoline{BFI}{Consc} &
\twoline{BFI}{Extra} & FAQ & IQCODE & \twoline{BFI}{Neuro} &
\twoline{BFI}{Open} & LC \\
\midrule

\multicolumn{11}{@{}l}{\textbf{Frozen  Embeddings + RF Regressor}} \\
Frozen + RF
& bert-large
& 0.108 & -0.006 & -0.035 & 0.023
& -0.041 & 0.086 & -0.001 & -0.010 & \textbf{0.057} \\

\midrule
\multicolumn{11}{@{}l}{\textbf{Frozen  Embeddings + Linear Regression Head}} \\
Frozen + Linear
& BioClinical-ModernBERT-base
& 0.086 & 0.028 & \textbf{0.209} & -0.080
& -0.084 & 0.188 & 0.141 & -0.256 & -0.201 \\

Frozen + Linear
& BioClinical-ModernBERT-large
& 0.028 & \textbf{0.073} & 0.079 & -0.138
& \textbf{0.174} & 0.059 & 0.173 & -0.022 & -0.064 \\

Frozen + Linear
& Clinical-Longformer
& \textbf{0.183} & -0.050 & -0.009 & -0.022
& 0.063 & \textbf{0.189} & 0.187 & -0.063 & -0.034 \\

Frozen + Linear
& distilbert-base-uncased
& 0.093 & -0.198 & 0.076 & \textbf{0.094}
& -0.039 & 0.166 & \textbf{0.218} & -0.132 & -0.135 \\

\midrule
\multicolumn{11}{@{}l}{\textbf{Baselines}} \\
Traditional Reg.
& linear\_regression
& 0.030 & 0.029 & 0.005 & 0.007
& -0.105 & 0.101 & 0.032 & \textbf{0.001} & 0.034 \\

Fine-tuning
& bert-large
& -0.042 & -0.132 & -0.119 & -0.219
& 0.070 & 0.014 & -0.072 & -0.008 & -0.097 \\

\bottomrule
\end{tabular}
\end{adjustbox}
\caption{Selected all-Q/A regression results for continuous cognitive, functional, and personality-related outcomes. Positive $R^2$ values indicate improvement over a mean-prediction baseline.}
\label{tab:results-aim3}
\end{table*}

\vspace{-1cm}

\subsection{Question-level Q/A Analysis for Perceived Cognitive Decline Classification}

Table~\ref{tab:question-level-results} presents an exploratory Single Q/A analysis. Several higher-ranked prompts directly ask about perceived cognitive or functional changes, including conversational attention (Q41), orientation in familiar areas (Q33), changes in social engagement (Q24), and driving difficulties (Q15). Because the primary target reflects participant-reported perceived cognitive decline, performance on such questions may partly reflect semantic overlap with the outcome.

Several indirect autobiographical or narrative questions also produced comparatively high test-set metrics. For example, Q08 asks participants to describe their previous job and daily responsibilities and yielded 0.8 accuracy, balanced accuracy, and macro-$F_1$. However, only 10 held-out participants had usable responses for Q08. Other questions used substantially larger test samples, including 31 participants for Q04, Q33, Q24, and Q37. Therefore, raw performance values across questions are not directly comparable without considering the effective sample size.

The question-level findings are exploratory for two additional reasons. First, for each question, the reported classifier was selected based on its performance on the same held-out test data. Second, the questions themselves were subsequently ranked using those same test results. This double use of the held-out set can produce optimistic estimates for the highest-ranked questions. Accordingly, the direct-versus-indirect comparison is hypothesis-generating rather than confirmatory. Future work should define the question categories a priori, select candidate questions and classifiers using only development data, and evaluate the resulting fixed subset once on an independent cohort.

\begin{table}[htbp]
\centering
\scriptsize
\setlength{\tabcolsep}{3pt}
\begin{adjustbox}{max width=\textwidth}
\begin{tabular}{p{0.05\linewidth} p{0.46\linewidth} l r r r r}
\toprule
QID & Question Text & Model & $n$ & Acc. & Bal. Acc. & Macro F1 \\
\midrule

Q41 & Have you noticed any changes in your ability to stay on track in conversations?
& SVM & 17 & \textbf{0.824} & 0.700 & 0.730 \\

Q08 & Think back to the last job you had. Can you describe your role and what your day-to-day responsibilities looked like?
& SVM & 10 & 0.800 & \textbf{0.800} & \textbf{0.800} \\

Q04 & What career path did you choose and why?
& DT & 31 & 0.742 & 0.698 & 0.705 \\

Q33 & Do you ever find yourself getting turned around in areas that you used to know well?
& RF & 31 & 0.742 & 0.698 & 0.705 \\

Q24 & Have you noticed a change in the past year with how much time you have spent with your friends or family?
& RF & 31 & 0.710 & 0.652 & 0.659 \\

Q40 & Can you describe your engagement during long conversations with friends or family?
& RF & 17 & 0.706 & 0.617 & 0.622 \\

Q18 & What was the wedding like? What did you wear?
& LR & 29 & 0.690 & 0.621 & 0.626 \\

Q37 & Finally, I would like you to tell me a short story. It can be about anything you want!
& RF & 31 & 0.677 & 0.607 & 0.609 \\

Q15 & In the past year, how has your driving experience been? Have you encountered any problems?
& LR & 28 & 0.643 & 0.633 & 0.626 \\

Q17 & Have you ever been married?
& LR & 14 & 0.643 & 0.646 & 0.641 \\

\bottomrule
\end{tabular}
\end{adjustbox}
\caption{Exploratory individual-question classification results. The effective held-out sample size varies across questions because not all questions were applicable to or answered by every participant. For each question, the displayed classifier was selected using performance on the same held-out set, and the questions were subsequently ranked using that set; therefore, the results may be optimistic and should be interpreted as hypothesis-generating.}
\label{tab:question-level-results}
\end{table}

\section{Limitations}

This study has several limitations. First, the modest sample ($N=153$; test $n=31$) yields wide participant-level bootstrap confidence intervals, limiting fine-grained comparisons among models. Second, the target reflects self-reported perceived cognitive decline rather than a clinically adjudicated diagnosis, and some ICA questions directly query cognitive or functional difficulties, creating potential semantic overlap with the outcome. Structured measures such as AD8 and IQCODE are also conceptually related to perceived cognitive change and should therefore be viewed as contextual reference measures rather than fully independent predictors. Third, the highest LLM configuration was identified descriptively among 18 test-set evaluations, while question-level analyses used variable sample sizes ($n=10$--31) and test-set-based model selection; these findings should therefore be considered exploratory. Finally, participants were recruited through Prolific, and proprietary LLMs may change over time. Larger, clinically characterized, and longitudinal cohorts are needed to assess generalizability and clinical utility.

\section{Conclusions and Future Work}

This study examined whether open-ended ICA chatbot interviews contain measurable signals associated with participant-reported perceived cognitive decline and related outcomes. LLM reasoning produced the numerically highest observed text-based classification results, although participant-level bootstrap confidence intervals overlapped substantially across the leading text-based models, and the highest configuration was selected descriptively among multiple test-set evaluations. Structured questionnaire, cognitive-task, and personality measures produced higher observed classification performance, while ICA text showed a modest predictive signal for several continuous outcomes. Exploratory question-level analyses further suggest that both direct symptom-related and indirect autobiographical or narrative prompts may contain useful information; however, variable sample sizes and test-set-based selection prevent confirmatory conclusions about individual questions. Overall, ICA interviews show promise for scalable research on perceived cognitive decline but do not establish clinical MCI detection or diagnosis. Future work should evaluate fixed models and question sets in larger, clinically characterized, longitudinal cohorts and investigate whether combining conversational text with speech, timing, structured assessments, and other multimodal signals improves generalizability.

\section{Acknowledgments}
\small
This research was partially supported by a GRIPex award from Kansas State University. This research was also partially sponsored by the Cognitive and Neurobiological Approaches to Plasticity (CNAP) Center of Biomedical Research Excellence (COBRE) of the National Institutes of Health (NIH) under grant number P20GM113109. The content is solely the responsibility of the authors and does not necessarily represent the official views of NIH.

\bibliography{ref}

@unpublished{prototype,
  title     = {Measuring Cognitive Functioning in Older Adults Using Artificial Intelligence Chatbot Narrative Interviews},
  author    = {Sun, Tianjun and Sylvara, Ashley and Heron, Xavier and Seitz, Christine and Caragea, Doina and Bailey, Heather},
  note      = {PsyArXiv Preprint},
  year      = {2025},
  publisher = {OSF},
  doi = {10.31234/osf.io/97tkx_v1},
  url ={https://doi.org/10.31234/osf.io/97tkx_v1}
}

@article{petersen1999mci,
  title={Mild cognitive impairment: Clinical characterization and outcome},
  author={Petersen, Ronald C. and Smith, Glenn E. and Waring, Susan C. and Ivnik, Robert J. and Tangalos, Eric G. and Kokmen, Emre},
  journal={Archives of Neurology},
  volume={56},
  number={3},
  pages={303--308},
  year={1999},
}

@article{petersen2018practice,
title={Practice guideline update summary: Mild cognitive impairment},
author={Petersen, Ronald C. and Lopez, Oscar and Armstrong, Melissa J. and Getchius, Thomas S. D. and Ganguli, Mary and Gloss, David and Gronseth, Gary S. and Marson, Daniel and Pringsheim, Tamara and Day, Gregory S. and others},
journal={Neurology},
volume={90},
number={3},
pages={126--135},
year={2018},
}

@article{nasreddine2005moca,
title={The Montreal Cognitive Assessment, {M}o{CA}: A brief screening tool for mild cognitive impairment},
author={Nasreddine, Ziad S. and Phillips, Natalie A. and B{\'e}dirian, Val{\'e}rie and Charbonneau, Simon and Whitehead, Victor and Collin, Isabelle and Cummings, Jeffrey L. and Chertkow, Howard},
journal={Journal of the American Geriatrics Society},
volume={53},
number={4},
pages={695--699},
year={2005},
}

@article{edmonds2019early,
title={Early versus late {MCI}: Improved {MCI} staging using a neuropsychological approach},
author={Edmonds, Emily C. and McDonald, Carrie R. and Marshall, Amanda and Thomas, Kelsey R. and Eppig, Joel and Weigand, Alexandra J. and Delano-Wood, Lisa and Galasko, Douglas R. and Salmon, David P. and Bondi, Mark W.},
journal={Alzheimer's \& Dementia},
volume={15},
number={5},
pages={699--708},
year={2019},
}

@article{aramaki2016vocabulary,
title={Vocabulary size in speech may be an early indicator of cognitive impairment},
author={Aramaki, Eiji and Shikata, Shoko and Miyabe, Mai and Kinoshita, Atsuko},
journal={PLOS ONE},
volume={11},
number={5},
pages={e0155195},
year={2016},
}

@article{fleming2014early,
title={Early detection of cognitive-linguistic change associated with mild cognitive impairment},
author={Fleming, Valerie B.},
journal={Communication Disorders Quarterly},
volume={35},
number={3},
pages={146--157},
year={2014},
}

@inproceedings{devlin2019bert,
title={BERT: Pre-training of deep bidirectional transformers for language understanding},
author={Devlin, Jacob and Chang, Ming-Wei and Lee, Kenton and Toutanova, Kristina},
booktitle={Proceedings of the 2019 Conference of the North American Chapter of the Association for Computational Linguistics},
pages={4171--4186},
year={2019},
}

@article{gemini,
  title={Introducing Gemini: our largest and most capable AI model},
  author={Pichai, Sundar and Hassabis, Demis},
  journal={Google. Retrieved December},
  volume={8},
  pages={2023},
  year={2023}
}

@article{gpt3,
  title={Language models are few-shot learners},
  author={Brown, Tom and Mann, Benjamin and Ryder, Nick and Subbiah, Melanie and Kaplan, Jared D and Dhariwal, Prafulla and Neelakantan, Arvind and Shyam, Pranav and Sastry, Girish and Askell, Amanda and others},
  journal={Advances in neural information processing systems},
  volume={33},
  pages={1877--1901},
  year={2020}
}

@article{taler2008language,
title={Language performance in Alzheimer's disease and mild cognitive impairment: A comparative review},
author={Taler, Vanessa and Phillips, Natalie A.},
journal={Journal of Clinical and Experimental Neuropsychology},
volume={30},
number={5},
pages={501--556},
year={2008},
doi={10.1080/13803390701550128}
}

@article{mueller2018connected,
title={Connected speech and language in mild cognitive impairment and Alzheimer's disease: A review of picture description tasks},
author={Mueller, Kimberly D. and Hermann, Bruce and Mecollari, Jola and Turkstra, Lyn S.},
journal={Journal of Clinical and Experimental Neuropsychology},
volume={40},
number={9},
pages={917--939},
year={2018},
doi={10.1080/13803395.2018.1446513}
}

@article{fan2023chatbot,
title={How well can an AI chatbot infer personality? {E}xamining psychometric properties of machine-inferred personality scores},
author={Fan, Jinyan and Sun, Tianjun and Liu, Jing and Zhao, Tian and Zhang, Bo and Chen, Zhen and Glorioso, Megan and Hack, Emily},
journal={Journal of Applied Psychology},
volume={108},
number={8},
pages={1277--1299},
year={2023}
}

@misc{claude,
  title        = {Introducing the Next Generation of Claude},
  author       = {{Anthropic}},
  year         = {2024},
  howpublished = {\url{https://www.anthropic.com/news/claude-3-family}},
  note         = {Accessed: 2026-05-27}
}

@inproceedings{optuna,
  title={Optuna: A next-generation hyperparameter optimization framework},
  author={Akiba, Takuya and Sano, Shotaro and Yanase, Toshihiko and Ohta, Takeru and Koyama, Masanori},
  booktitle={Proceedings of the 25th ACM SIGKDD international conference on knowledge discovery \& data mining},
  pages={2623--2631},
  year={2019}
}

@article{langa2014diagnosis,
  title={The diagnosis and management of mild cognitive impairment: a clinical review},
  author={Langa, Kenneth M and Levine, Deborah A},
  journal={Jama},
  volume={312},
  number={23},
  pages={2551--2561},
  year={2014}
}

@article{apostolo2016predicting,
  title={Predicting risk and outcomes for frail older adults: results of an umbrella review of available screening tools for frailty},
  author={Ap{\'o}stolo, J and Bobrowicz-Campos, E and Holland, C and Cooke, R and Santana, S and Marcucci, M and Vollenbroek, M and Germini, F and Cano, A},
  journal={JBI Evidence Implementation},
  volume={14},
  number={4},
  pages={192},
  year={2016},
  publisher={LWW}
}

@article{bucks1,
  title={Analysis of spontaneous, conversational speech in dementia of Alzheimer type: Evaluation of an objective technique for analysing lexical performance},
  author={Bucks, Romola S and Singh, Sameer and Cuerden, Joanne M and Wilcock, Gordon K},
  journal={Aphasiology},
  volume={14},
  number={1},
  pages={71--91},
  year={2000},
  publisher={Taylor \& Francis}
}

@article{bucks2,
  title={Disentangling the underlying mechanisms of verbal fluency decline in cognitive aging.},
  author={Kljajevic, Vanja},
  journal={Psychology \& Neuroscience},
  volume={19},
  number={1},
  pages={42},
  year={2026},
  publisher={Educational Publishing Foundation}
}

@article{bucks3,
  title={Dementia detection using automatic analysis of conversations},
  author={Mirheidari, Bahman and Blackburn, Daniel and Walker, Traci and Reuber, Markus and Christensen, Heidi},
  journal={Computer Speech \& Language},
  volume={53},
  pages={65--79},
  year={2019},
  publisher={Elsevier}
}

@misc{sun2024development,
  title={Development and validation of an artificial intelligence chatbot to assess personality},
  author={Sun, Tianjun and Roberts, Brent and Drasgow, Fritz and Zhou, Michelle X},
  year={2024},
  publisher={OSF}
}

@article{soto2017next,
  title={The next Big Five Inventory (BFI-2): Developing and assessing a hierarchical model with 15 facets to enhance bandwidth, fidelity, and predictive power.},
  author={Soto, Christopher J and John, Oliver P},
  journal={Journal of personality and social psychology},
  volume={113},
  number={1},
  pages={117},
  year={2017},
  publisher={American Psychological Association}
}
\end{document}